\documentclass[amsmath,showpacs,aps,twocolumn,prb]{revtex4-1}

\usepackage{bm}
\usepackage{graphics}
\usepackage{graphicx}
\usepackage[dvips]{epsfig}

\begin{document}
\title{Coulomb drag between helical edge states}
\author{Vladimir A. Zyuzin}
\author{Gregory A. Fiete}
\affiliation{Department of Physics, The University of Texas at Austin, Austin, TX 78712, USA}
\begin{abstract}
We theoretically investigate the Coulomb drag between the edge states of two quantum spin Hall systems. Using an interacting theory of the one-dimensional helical edge modes, we show that the drag vanishes at second order in the inter-edge interaction, where it is typically finite in other systems, due to the absence of backscattering within the edges.  However, in the presence of a small external magnetic field the drag is finite and scales as the fourth power of the magnetic field, a behavior that sharply distinguishes it from other systems.  We obtain the temperature dependence of the drag for regimes of both linear and quadratic edge dispersion in the presence of a finite field.
\end{abstract}
\maketitle

{\it Introduction--}Topological phases of matter have attracted interest because their quantum properties are robust to many material imperfections.\cite{Volovik,Wen,Nayak:rmp08}  In particular, the quantum spin Hall (QSH) system has recently emerged\cite{Kane:prl05,Kane_2:prl05,Bernevig:prl06} as a time-reversal invariant counterpart to the integer quantum Hall effect.\cite{Hasan:rmp10,Moore:nat10,Qi:pt10} Shortly after the prediction\cite{Bernevig:sci06} that the HgTe/(Hg,Cd)Te quantum well system should exhibit a QSH state, the experimental observation was made\cite{Konig:sci07} and further confirmation followed.\cite{Roth:sci09}

The QSH state has an insulating bulk and metallic edge states composed of an odd number of Kramer's pairs of electrons.  A $Z_2$ invariant distinguishes the topological insulators with time-reversal symmetry from their ``trivial" counterparts.\cite{Kane:prl05,Kane_2:prl05,Bernevig:prl06}  The simplest topologically non-trivial case is a single Kramer's pair on the edge.   Due to the spin-orbit coupling that drives the QSH state, the spin of an electron on the edge is correlated with its momentum.  This property leads to an absence of back-scattering from weak non-magnetic impurities and therefore prevents Anderson localization on the edge of the QSH system.\cite{Wu:prl06,Xu:prb06}

The gapless edge modes of the QSH system are commonly referred to as a helical liquid (HL).\cite{Wu:prl06}  The stability of the HL to interactions,\cite{Wu:prl06,Xu:prb06} and magnetic disorder\cite{Wu:prl06,Maciejko:prl09} has been investigated, as has its response to ``pinching" the sample into a point contact\cite{Strom:prl09,Law:prb10,Teo:prb09} or related geometries.\cite{Huo:prl09}  Properties of superconducting-QSH hybrid structures were investigated as well.\cite{Xu:prb10}  When two HL (of different QSH systems) are allowed to interact with each other, a novel one-dimensional correlated state is formed at the lowest energies.\cite{Tanaka:prl09}

\begin{figure}[th]
\includegraphics[width=5cm]{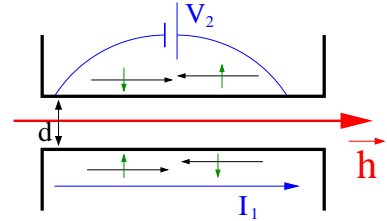}
\caption{(color online) Schematic of a drag measurement between two QSH systems.  A current $I_1$ is driven along the upper edge of the lower QSH system and through electron-electron interactions a voltage $V_2$ is induced in the lower edge of the upper QSH system. A magnetic field $\vec h$ is applied in the plane of wires, perpendicular to the spin quantization axis (assumed perpendicular to the plane of QSH systems).  Time-reversed Kramer's pairs are indicated for the two edges.  A QSH on top of QSH geometry could also be used.} \label{fig1}
\end{figure}

In this paper we study the Coulomb drag between two QSH systems, each having one Kramer's pair as shown schematically in Fig.\ref{fig1}.  The drag experiment we discuss should be carried out at energy (temperature) scales above which the inter-edge correlated state forms,\cite{Tanaka:prl09} but below the bulk energy gap of the QSH system.  In a Coulomb drag experiment current is driven in an ``active" wire/edge and voltage is measured in a ``passive" wire/edge.  The Coulomb interaction between electrons in different system results in a momentum transfer between the two systems and produces a voltage drop in the ``passive" system. The drag is characterized by the drag resistivity,
\begin{equation}
r_{D}=-\lim_{I_{1}\rightarrow 0}\frac{e^2}{h}\frac{1}{L}\frac{dV_{2}}{dI_{1}},
\end{equation}
where $V_{2}$ is voltage induced in the ``passive" system by the current $I_{1}$ driven in ``active" system. Here $e$ is the electron charge, $h$ is a Planck's constant, and $L$ is the length of the edge along which momentum is transferred.

Coulomb drag  in non-HL one-dimensional systems has been studied both theoretically\cite{Klesse:prb00,Ponomarenko:prl00,Fuchs:prb05,Pustilnik:prl03,Aristov:prb07,Fiete:prb06} and experimentally.\cite{Yamamoto:pe02,Yamamoto:Sci06,Debray:sst01,Debray:jpcm01} The HL can be viewed as a spinless Luttinger liquid (because it has the same number of degrees of freedom) without backscattering; since it is known that the backscattering governs the drag between systems with linear dispersions,\cite{Klesse:prb00,Pustilnik:prl03,Fiete:prb06} one can not expect drag between helical liquids. However, in this paper we show that an applied Zeeman field $\vec h$ opens up a backscattering process in the HL, and results in $r_D \propto h^4$ for a linear spectrum. We also compute the temperature dependence of the drag over a range of temperature and field values.

{\it The model--}In second order perturbation theory in the interwire Coulomb interaction, the Coulomb drag is given by\cite{Pustilnik:prl03}
\begin{equation}
\label{eq:r_D}
r_{D}=\int_{0}^{\infty}dq~\int_{0}^{\infty}d\omega ~ \frac{q^2U^{2}_{12}(q)}{4\pi^{3}n_{1}n_{2}T}\frac{\Im \Pi_{1}(q,\omega)\Im \Pi_{2}(q,\omega)}{\sinh^2(\frac{\omega}{2T})},
\end{equation}
where $\Im \Pi_{i}(\omega,q)$ is the imaginary part of the retarded density-density correlation function of wire $i=1(2)$, $n_{i}$ is electron density of wire $i$, $T$ is the temperature, and $U_{12}(q)$ is the Fourier transform of the interwire Coulomb interaction which is cut off at short distances by the interwire separation $d$.

\begin{figure}
\includegraphics[width=6cm]{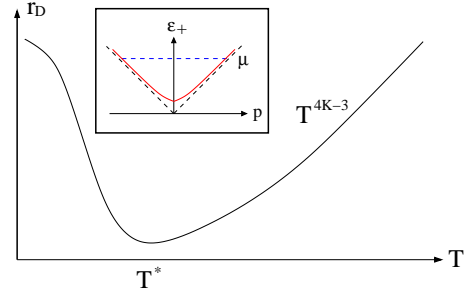}
\caption{(color online) Temperature dependence of the drag in the regime of small $h$ where the spectrum may be approximated as linear, as shown in the inset. $T^*$ is the temperature at which the wires begin to ``lock" to each other.\cite{Klesse:prb00}  For $T >T^*$, we find $r_D \propto h^4 T^{4K-3}$ where $K$ is the Luttinger parameter in the charge sector.} \label{fig2}
\end{figure}

\begin{figure}
\begin{center}
\includegraphics[width=6cm]{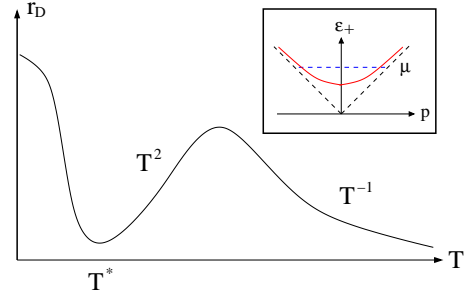}
\caption{ (color online) Temperature dependence of the drag in the regime of small $\mu$ where the spectrum is approximately quadratic, as shown in the inset, and $0<\mu-h < \frac{v^2}{2d^2 h}$. Note the non-monotonic temperature dependence\cite{Fiete:prb06} above $T^*$.  For the dependence of $r_D$ on the Zeeman field $h$ in each region of temperature, see the text. The second crossover from $T^2$ to $T^{-1}$ occurs for $T\sim  \frac{v}{4d}\sqrt{\frac{\mu - h}{2h}}$ where $d$ is the distance between wires. } \label{fig3}
\end{center}
\end{figure}

We consider two identical QSH systems, each with one Kramer's pair on its edge, as shown in Fig.\ref{fig1}. As we noted earlier, if the spectrum is linear there is no contribution to the drag from forward scattering, and back scattering is forbidden by time-reversal symmetry.  Therefore, one must break time-reversal symmetry in order to open up a backward scattering channel (unless there are magnetic impurities present) for a generic Dirac edge mode.  Our Hamiltonian for a single HL in the presence of a Zeeman field is
\begin{equation}
H_{0}=\int dx {\hat \psi}^{\dag}(x) \left( v{\hat p}_{x} {\hat \sigma}_{z} + h{\hat \sigma}_{x} -\mu\right){\hat \psi}(x),
\end{equation}
where $v$ is the edge velocity, ${\hat p}_{x}=-i\partial_{x}$, $\mu$ is the Fermi energy (which can be adjusted by gating the system), and $\hat \sigma_{z,x}$ are Pauli spin matricies describing the spin degree of freedom. A Zeeman field $\vec h$ pointing in the $x$-direction opens up a gap in Dirac spectrum and tilts the spins away from the $z$-axis. The edge dispersion is $\epsilon_{\pm}=\pm\sqrt{v^2p^2+h^2}-\mu$. We assume that Fermi energy is in the upper band ($\mu>0$) so that the properties of system are determined by the $\epsilon_{+}$ band over the energy scales of interest. The wavefunction of electrons in the $\epsilon_{+}$ band is
\begin{equation}
\label{eq:psi}
{\hat \psi}_{+}(x)= \frac{e^{ipx}}{\sqrt{2}}\left(\begin{array}{c}
\cos(\gamma_{p}/2) + \sin(\gamma_{p}/2)\\
\frac{\cos(\gamma_{p})}{(\cos(\gamma_{p}/2) + \sin(\gamma_{p}/2))}\\
\end{array} \right)=e^{ipx}{\hat U}_{p},
\end{equation}
where $\gamma_{p}=\arctan(\frac{vp}{h})$. We study \eqref{eq:psi} in the limit of large $\mu$ (small $h$) where the spectrum can be approximated as linear, and also in the opposite limit where the spectrum is approximately quadratic ({\it i.e.}, $\mu$ close to the band ``bottom").  See Figs. \ref{fig2} and \ref{fig3}.

{\it Regime of linear spectrum--}We first consider the case $\mu-h>h$, and linearize the spectrum near the Fermi energy, $\epsilon_{+}=v|p_{x}|-\mu$ (see Fig. \ref{fig2}.), in order to use standard bosonization procedures.\cite{Giamarchi}  We express the electron operator as a sum of left- and right- moving states: ${\hat \psi}_{+}={\hat \psi}_{R}(x)+{\hat \psi}_{L}(x)$, where $R(L)$ stands for right (left) movers. The non-interacting Hamiltonian can then be written:
\begin{equation}
H_{0}=\int dx\Bigl[{\hat \psi^{\dag}}_{R}(x)\hat p_+{\hat \psi}_{R}(x)
+ {\hat \psi}^{\dag}_{L}(x)\hat p_-{\hat \psi}_{L}(x)\Bigr].
\end{equation}
where $\hat p_\pm=\pm v \hat p_x -\mu$. We assume intrawire interactions have the form
\begin{equation}
\label{eq:H_int}
H_{int}=\frac{1}{2}\int dxdx'~U(x-x')\rho(x)\rho(x'),
\end{equation}
where $U(x-x')$ is the intrawire Coulomb interaction, and $\rho(x)$ is the electron density:
\begin{equation}
\rho(x)={\hat \psi}_{R}^{\dag}{\hat \psi}_{R}+{\hat \psi}_{L}^{\dag}{\hat \psi}_{L}+\cos(\gamma_{p})\left({\hat \psi}_{R}^{\dag}{\hat \psi}_{L}+{\hat \psi}_{L}^{\dag}{\hat \psi}_{R}\right),
\end{equation}
which contains cross terms due to the presence of the magnetic field. In terms of bosonic fields $\phi(x)$ and $\theta(x)$,  ${\hat \psi}_{R}$ and ${\hat \psi}_{L}$ are expressed as\cite{Giamarchi}
\begin{eqnarray}
&{\hat \psi}_{R}(x)=e^{ipx}\frac{\eta_{R}}{\sqrt{2\pi a}}e^{-i(\phi(x)-\theta(x))}, \\
&{\hat \psi}_{L}(x)=e^{-ipx}\frac{\eta_{L}}{\sqrt{2\pi a}}e^{-i(-\phi(x)-\theta(x))},
\end{eqnarray}
where $\eta_{R(L)}$ are Klein factors, and $a$ is a short-distance cut-off.  The electron density in terms of bosonic fields takes the form
\begin{equation}\label{density}
\rho(x)=-\frac{1}{\pi}\partial_{x}\phi(x)-\frac{\cos(\gamma_{p_{F}})}{\pi a}\sin(2p_{F}x-2\phi(x)).
\end{equation}
Substituting this expression into \eqref{eq:H_int}, we find
\begin{equation}\label{interaction}
H_{int}=\frac{U(0)-\cos^{2}(\gamma_{p_{F}})U(2p_{F})}{2\pi^2}\int dx~ (\partial_{x}\phi(x))^2,
\end{equation}
where $U(0)$ and $U(2p_{F})$ are the zero and $2p_{F}$ momentum parts of the interaction, respectively. Note that the $2p_{F}$ part has a $\cos^{2}(\gamma_{p_{F}})$ factor, which is proportional to $h^2$ for small $h$. The full Hamiltonian then becomes
\begin{equation}\label{ham_mag_linear}
H=\frac{1}{2\pi}\int dx~\left[v(\partial_{x}\theta(x))^2 + \left(v + g \right)  (\partial_{x}\phi(x))^2\right],
\end{equation}
where $g=(U(0)-\cos^{2}(\gamma_{p_{F}})U(2p_{F}))/\pi$. We observe that the Hamiltonian of an interacting HL in a Zeeman field is equivalent to a spinless Luttinger liquid where the strength of backscattering depends on the Zeeman  field. Similar results were obtained in studies of Luttinger liquids with Rashba spin-orbit coupling and a Zeeman magnetic field.\cite{Sun:prl07,Gangadharaigh:prb08}

We now give an expression for the retarded density-density correlation function. Because the backscattering governs the Coulomb drag when the dispersion is linear (as it is in a Luttinger liquid), we only need the $2p_{F}$ part of the retarded density-density correlation function. Since our model has reduced to a spinless Luttinger liquid, the calculation is standard,\cite{Giamarchi}
\begin{equation}
\Pi^{2p_{F}}_{R}(q,\omega)=\frac{1}{2}\left[{\tilde \Pi}(q+2p_{F},\omega) + {\tilde \Pi}(q-2p_{F},\omega)\right],
\end{equation}
with ${\tilde \Pi}(q,\omega)$ is given by
\begin{equation}
\label{eq:Pi}
{\tilde \Pi}(q,\omega)=\frac{2^{2K}D}{u}\left(\frac{\beta u}{2\pi} \right)^{2}F(q,\omega),
\end{equation}
where $F(q,\omega)=B(-i\frac{\beta }{4\pi}(\omega - uq)+\frac{K}{2},1-K)B(-i\frac{\beta }{4\pi}(\omega + uq)+\frac{K}{2},1-K)$, $\beta=1/T$, $u=\sqrt{v(v+g)}$, $K=v/u$, and $B(x,y)$ is the Beta function. The parameter $D$ is
\begin{equation}
\label{eq:D}
D=\cos^{2}(\gamma_{p_{F}})\sin(\pi K)\frac{(\pi a)^{2K-2}}{(u\beta)^{2K}}.
\end{equation}

With  \eqref{eq:Pi} in hand, the drag resistivity is readily computed from \eqref{eq:r_D}: $\Im F(q\pm 2p_{F},\omega)$ is sharply peaked about $q=\mp 2p_{F}$ with peak widths proportional to temperature.  Since the momentum integration in \eqref{eq:r_D} runs from $0$ to $\infty$, we neglect the ${\tilde \Pi}(q+2p_{F},\omega)$ contribution. Taking the imaginary part of the retarded density-density correlation function and assuming identical helical liquids we obtain
\begin{equation}
r_{D}\simeq\frac{2^{4K}u^{2}D^{2}}{16^2\pi^{7}}(2p_{F})^{2}U_{12}^{2}(2p_{F})\frac{I}{n^{2} T^{3}},
\end{equation}
where $I \equiv \int_{0}^{\infty}d\Omega~\frac{(\Im F(0,\Omega))^2}{\sinh^{2}(\Omega/2)}$, with $\Omega=\omega/T$. The density of states, $n=1/\pi v$. Extracting the temperature and magnetic field dependence using \eqref{eq:D},  we find
\begin{equation}
\label{result1}
r_{D} \propto h^{4}T^{4K-3}.
\end{equation}
Eq.\eqref{result1} is one of the central results of the paper.  This result is valid at temperatures larger than $T^{*}$, below which the drag begins to exhibit an exponential dependence on temperature.\cite{Klesse:prb00,Ponomarenko:prl00}  Since $T^{*}\sim \mu e^{-\frac{p_F d}{1-K}}$ depends\cite{Klesse:prb00} on the backscattering via $K$, it will also depend on the Zeeman field via the dependence of $K$ on $h$.

By contrast, in a spinful Luttinger liquid the magnetic field only enters the interaction constant in the spin channel and therefore the drag is only (weakly) dependent on magnetic field through the interaction parameter appearing in an exponent to the temperature. Therefore, Coulomb drag can be used as a method for experimental verification of the HL, complementing the earlier studies.\cite{Wu:prl06,Maciejko:prl09,Strom:prl09,Law:prb10,Teo:prb09,Huo:prl09,Xu:prb10,Tanaka:prl09}
We note that a spin-Coulomb drag effect in which two density mismatched Luttinger liquids can be brought into more favorable kinematic conditions for enhanced drag effects has also been studied.\cite{Pustilnik:prl06}  To complete our analysis of drag between two HL, we turn to the case when the spectrum is approximately quadratic.

{\it Regime of quadratic spectrum--}When $0<\mu-h<h$, the spectrum of upper band is approximately  $\epsilon_{+}=\frac{1}{2}\frac{(vp)^2}{h}-(\mu - h)$, as seen in Fig. \ref{fig3}.  In this section, we simplify the problem by assuming no intrawire electron interactions. The neglect of weak interactions in the regime of a quadratic dispersion has been shown to have no effect on the temperature dependence of the drag.\cite{Pustilnik:prl03} The imaginary part of the retarded density-density correlation function is
\begin{equation}
\label{eq:Pi_quad}
\Im \Pi_{R}(q, \omega)=-\frac{h}{4v^2 q}f^{2}_{+}(p_{0},q) \frac{\sinh(\frac{\omega}{2T})}{\cosh(\frac{\epsilon_{+}(p_{0})}{2T})\cosh(\frac{\epsilon_{+}(p_{0}+q)}{2T})},
\end{equation}
where $p_{0}=-\frac{1}{2}q+\frac{h\omega}{v^2 q}$, and $f_{+}(p,q)={\hat U}^{\dag}_{p}{\hat U}_{p+q}$, which we assume to be approximately equal to one.  One also needs to take into account restrictions on $\omega$ defined by $\epsilon_{+}(p_{0})<0$ and $\epsilon_{+}(p_{0}+q)>0$ which will give
\begin{equation}
\frac{1}{2}vq + \sqrt{2h(\mu - h)} > \frac{h\omega}{vq} >-\frac{1}{2}vq + \sqrt{2h(\mu - h)}.
\end{equation}
Plugging \eqref{eq:Pi_quad} into \eqref{eq:r_D} and evaluating the integrals we obtain the following results. When $\mu-h > \frac{v^2}{2d^2 h}$ [here $d$ is the inter-edge separation distance and $1/d$ serves as high momentum cut-off to $U_{12}(q)$ in the $q$ integration in \eqref{eq:r_D}] and for small temperatures $T<\frac{v}{4d}\sqrt{\frac{\mu - h}{2h}}$,
\begin{equation}\label{result2}
r_{D}\simeq \frac{1}{2^5\pi^2 n^2}\frac{g_\gamma^2}{v^4}\sqrt{\frac{h^5}{(\mu-h)^3}}T^2,
\end{equation}
while at large temperatures $T>\frac{v}{4d}\sqrt{\frac{\mu - h}{2h}}$,
\begin{equation}\label{result3}
r_{D}\simeq \frac{1}{2^8 \pi^3 n^2}\frac{hg_\gamma^2}{vd^3}\frac{1}{T}.
\end{equation}
When $\mu-h < \frac{v^2}{2d^2 h}$ and for small temperatures $T<\frac{v}{d}\sqrt{\frac{\mu - h}{2h}}$,
\begin{equation}\label{result4}
r_{D}\simeq \frac{2^{7/2}}{\pi^2 n^2}\frac{g_\gamma^2}{v^4}h^{5/2}\sqrt{T},
\end{equation}
while at large temperatures $T>\frac{v}{d}\sqrt{\frac{\mu - h}{2h}}$
\begin{equation}\label{result5}
r_{D}\simeq\frac{2^{15/2}}{3\pi^3 n^2}\frac{g_\gamma^2}{v}\frac{h(\mu-h)^{3/2}}{d^3}\frac{1}{T^{5/2}}.
\end{equation}
Here $g_\gamma=-\gamma+\ln(2)$ ($\gamma \approx 0.5772$ is Euler's constant) is an estimate of interwire Coulomb interaction at small momentum.  The density of states, $n=\frac{1}{\pi}\frac{1}{\sqrt{\mu-h}}$. The results are summarized in Fig. \ref{fig3}.  We emphasize that in obtaining these results we have not considered effects of interband (intra-edge) particle-hole excitations. These excitations will result in Fermi edge singularity physics.\cite{Pustilnik_2:prl06,Fiete:jpcm09}

{\it Summary and Discussion--}We studied the Coulomb drag between identical one dimensional helical liquids. We showed that the helical liquid can be mapped to a spinless Luttinger liquid where backscattering is prohibited. Since backscattering governs the drag between one-dimensional liquids with linear dispersion, there is no Coulomb drag unless there is a nonlinearity in the spectrum. Nonlinearity in the spectrum gives rise to a small momentum scattering contribution to the drag which has a $T^2$ temperature dependence.\cite{Pustilnik:prl03} Our calculations confirm these results for a nonlinear spectrum.

For a linear spectrum, the application of a Zeeman field opens up backscattering processes which are proportional to the square of the magnetic field at small fields, see (\ref{density}) and (\ref{interaction}). We also showed that when the magnetic field is small ($\mu$ large) and the spectrum can be approximated as linear, the Hamiltonian of a helical liquid with a magnetic field is identical to that  of a spinless Luttinger liquid with a magnetic field dependent backscattering term (\ref{ham_mag_linear}). In this case, the Coulomb drag becomes proportional to the fourth power of magnetic field (\ref{result1}) which is distinct from the case of Luttinger liquids where the magnetic field enters only via the interaction constant in the spin channel.

For completeness we studied the case when the spectrum of a helical liquid in a magnetic field can not be approximated as linear, but is rather approximately quadratic (valid for a strong magnetic field). Expressions for the Coulomb drag in this case are given by (\ref{result2})-(\ref{result5}). Finally, we note that the presence of a few magnetic impurities on the edge of a QSH system would allow a finite drag contribution even in the absence of applied magnetic fields since they would allow backscattering.  Inclusion of Rashba coupling would not affect our results, provided the zero-field case is still adiabatically connected to the topologically non-trivial state.

We are grateful to Jun Wen and Suhas Gangadharaiah for stimulating discussions and ARO grant W911NF-09-1-0527 for financial support.


%

\end{document}